\documentclass[aps,prb,showpacs,twocolumn,groupedaddress]{revtex4}

\usepackage[english]{babel}
\usepackage{epsfig}
\usepackage{latexsym}

\begin{document}
\title{\emph{Multiple} double-exchange mechanism by Mn$^{2+}$-doping in manganite compounds}

\author{P.Orgiani$^{1,2}$, A.Galdi$^{1,3}$, C.Aruta$^{4}$, V.Cataudella$^{4}$,
G.De Filippis$^{4}$, C.A.Perroni$^{4}$, V.Marigliano
Ramaglia$^{4}$, R.Ciancio$^{5}$, N.B.Brookes$^{6}$, M.Moretti
Sala$^{7}$, G.Ghiringhelli$^{7}$, and L.Maritato$^{1,2}$}

\address{\scriptsize{$^{1}$CNR-SPIN, I-84084 Fisciano (SA), Italy.}}
\address{\scriptsize{$^{2}$Department of Mathematics and Informatics, University of Salerno, I-84084 Fisciano (SA), Italy.}}
\address{\scriptsize{$^{3}$Department of Physics, University of Salerno, I-84084 Fisciano (SA), Italy.}}
\address{\scriptsize{$^{4}$CNR-SPIN and Department of Physics, University of Napoli, I-80126 Napoli, Italy.}}
\address{\scriptsize{$^{5}$CNR-IOM, TASC Laboratory, I-34149 Trieste, Italy.}}
\address{\scriptsize{$^{6}$European Synchrotron Radiation Facility, F-38043 Grenoble, France.}}
\address{\scriptsize{$^{7}$CNR-SPIN and Department of Physics, Politecnico di Milano, I-20133 Milano, Italy.}}

\date{\today}
\begin{abstract}

\scriptsize{Double-exchange mechanisms in
RE$_{1-x}$AE$_{x}$MnO$_{3}$ manganites (where RE is a trivalent
rare-earth ion and AE is a divalent alkali-earth ion) relies on
the strong exchange interaction between two Mn$^{3+}$ and
Mn$^{4+}$ ions through interfiling oxygen 2p states. Nevertheless,
the role of RE and AE ions has ever been considered "silent" with
respect to the DE conducting mechanisms. Here we show that a new
path for DE-mechanism is indeed possible by partially replacing
the RE-AE elements by Mn$^{2+}$-ions, in La-deficient
La$_{x}$MnO$_{3-\delta}$ thin films. X-ray absorption spectroscopy
demonstrated the relevant presence of Mn$^{2+}$ ions, which is
unambiguously proved to be substituted at La-site by Resonant
Inelastic X-ray Scattering. Mn$^{2+}$ is proved to be directly
correlated to the enhanced magneto-transport properties because of
an additional hopping mechanism trough interfiling Mn$^{2+}$-ions.
Such a scenario has been theoretically confirmed by calculations
within the effective single band model. The use of Mn$^{2+}$ both
as a doping element and an ions electronically involved in the
conduction mechanism reveals a new phenomena in transport
properties of manganites. More important, such a strategy might be
also pursed in other strongly correlated materials.}

\end{abstract}
\thanks{Corresponding author; email address: pasquale.orgiani@spin.cnr.it}
 \maketitle

\section{Introduction}

Since its discovery, colossal magneto-resistance (CMR) effect has
undoubtedly been among the most studied phenomena in solid state
physics\cite{cmr,salamon,coey,dagotto}. CMR phenomenon has been
explained within the framework of double exchange (DE) mechanism,
based on a strong exchange interaction between Mn$^{+3}$ and
Mn$^{+4}$ ions through intervening filled oxygen 2p
states\cite{de,good}. The metallicity in these systems, as its
strong dependence from an external magnetic field, comes from the
mixed valence of Mn ions, which can transfer both charge and spin
between their Mn$^{+3}$ and Mn$^{+4}$ states. Due to Hund's rule,
this transfer occurs only if the core spin of Mn$^{+3}$ is aligned
with that of Mn$^{+4}$, thus explaining the colossal effects on
the resistance values by the application of a spin-aligning
external magnetic field. CMR effect has been intensively
investigated in RE$_{1-x}$AE$_{x}$MnO$_{3}$ (where RE is a
trivalent rare-earth ion and AE is a divalent alkali-earth ion)
manganites. The mandatory Mn$^{+3}$/Mn$^{+4}$ mixed population is
generally controlled by the chemical substitution of the trivalent
RE$^{3}+$-ion with a divalent AE$^{2+}$-ion. Indeed, to satisfy
the overall charge neutrality within the manganite unit cell, when
RE$^{3+}$-ions are substituted by AE$^{2+}$, some of the Mn atoms
are forced into a 4+ state\cite{cmr, salamon, coey, dagotto,
good}. Nevertheless, with respect to the DE hopping mechanism, the
role of AE-RE elements has always been considered as "silent", by
being their corresponding conduction bands too far from Fermi's
energy level. In order to make such an atomic site active within
the transport mechanism, a possible strategy calls for a RE$^{3+}$
substitution by using multiple-valence Mn-ions themselves.

Here we show that a partial substitution of Mn-ions at the La-site
is indeed possible in La-deficient La$_{x}$MnO$_{3-\delta}$
manganite thin films. By combining polarization dependent x-ray
absorption spectroscopy (XAS) and resonant inelastic x-ray
spectroscopy (RIXS), the relevant Mn$^{2+}$ content is
demonstrated, and it is unambiguously assigned its
crystallographic site (namely, the La-site). Similarly to
AE$^{2+}$-doped manganites, the La$^{3+}$/Mn$^{2+}$ substitution
induces the required Mn$^{+3}$/Mn$^{+4}$ mixed population.
However, differently from the AE$^{2+}$-doping, the Mn$^{2+}$-ions
at La-site are electronically involved in the transport
mechanisms, having their electronic bands crossing the Fermi
energy. Such an energetic configuration favors the hopping of
electrical charge through that site (usually silent), in addition
to the traditional Mn$^{+3}$/Mn$^{+4}$ hopping path, thus
contributing to the ferromagnetic and metallic state. Such a
\emph{Multiple}-DE phenomenon has never been reported and it opens
new perspectives in both fundamental studies on transport
mechanism in strongly correlated manganites as in possible
application by using these new class of materials.

\section{Thin film growth and structural properties}

LMO thin films, with different values of the La/Mn ratio, were
fabricated by Molecular-Beam Epitaxy (MBE) on SrTiO$_{3}$ (STO)
substrates. Oxygen content was systematically varied by
postannealing all the LMO samples, by varying the duration and the
temperature of the process (thus providing LMO films at different
stages of oxygenation, i.e. oxygen content)\cite{apl2009}. X-ray
diffraction (XRD) analyses show that all the investigated LMO
samples are in-plane matched with STO substrates. However, the
out-of-plane lattice parameter monotonically depends on the La/Mn
stoichiometric ratio. For each annealing batch (i.e., oxygen
content), such a systematic trend in c-lattice parameter has
always been observed within the LMO samples with different La/Mn
stoichiometric ratio, which appears to be the major factor in
c-lattice variation within each set of samples. XRD spectra for
optimized LMO series (i.e., highest metallicity) are reported in
Fig.1.

\begin{figure}[!h]
\includegraphics[width=8.5 cm]{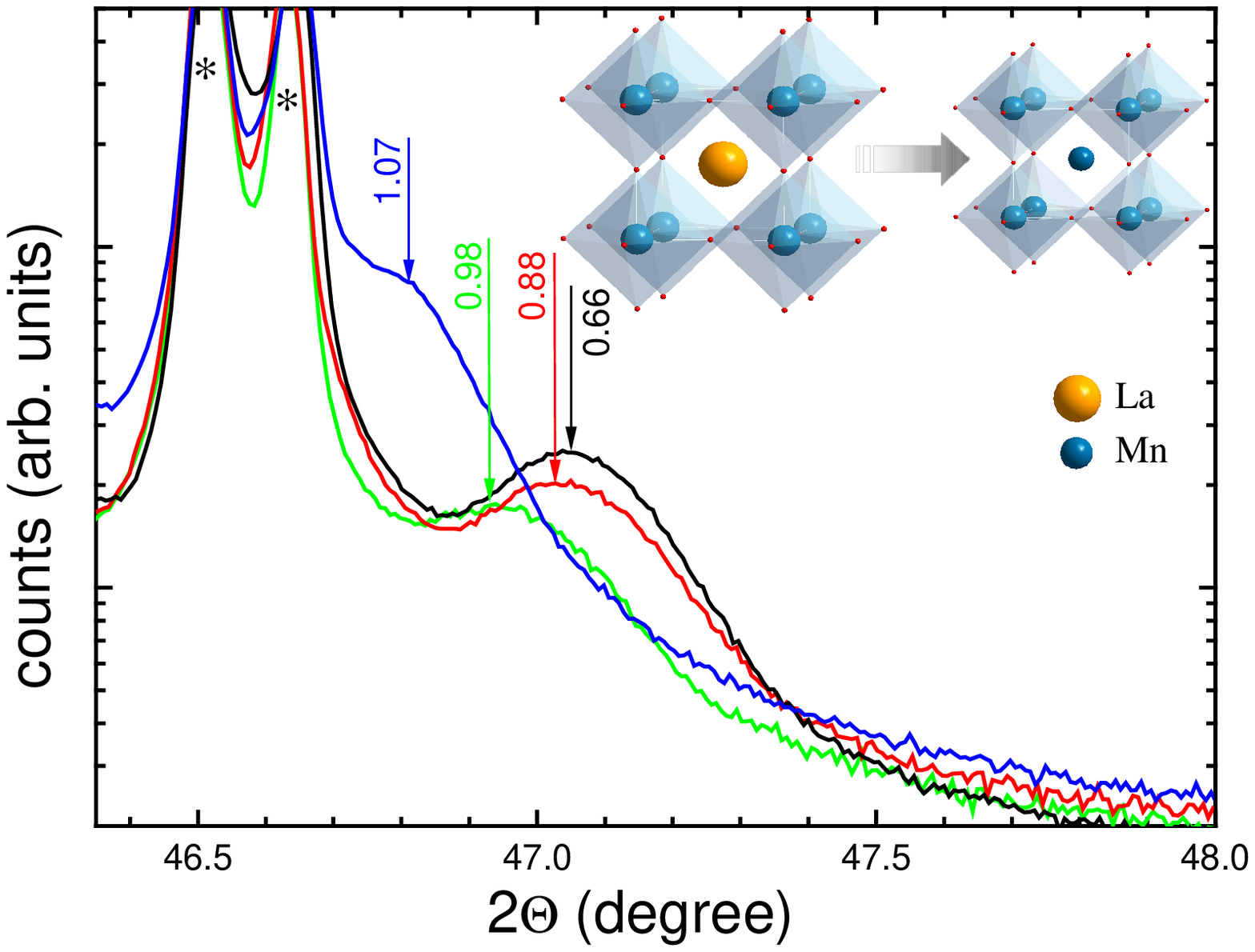}
\caption{\scriptsize{(Color online) $\theta$-2$\theta$ scan around
the (002) Bragg reflection of four LMO films with La/Mn ratio 0.66
(black), 0.88 (red), 0.98 (green) and 1.07 (blue), respectively
(asterisks indicate the STO peaks). In the inset, a cartoon
showing the expected reduction of the unit cell by the
Mn-substitution (blue spheres) at the La-site (orange sphere) is
also sketched.}}
\end{figure}

All the spectra only shows the [00l] peaks, indicating the
preferential orientation of the film along the [001] substrate
crystallographic direction, and no secondary phase is detected.
The out-of-plane lattice parameters vary from 3.859 (measured in
the sample with La/Mn = 0.66) to 3.878 \AA (sample with La/Mn =
1.07). Considering that Mn ionic radius (which varies from 0.53 to
0.89\AA, depending on its electronic configuration) is sizably
smaller than those measured for the La-ions (ranging from 1.17 to
1.5\AA), the monotonic decrease of the out-of-plane lattice
parameter values is compatible with the gradual substitution of Mn
atoms at the La-site, which was confirmed possible by neutron
diffraction investigations of bulk samples\cite{neutron}. It is
worthful to remark that, in polycrystalline LMO bulk-samples, it
was shown a La-vacancies limiting value of 0.125 (i.e.,
La$_{0.875}$MnO$_{3}$), while further excess of Mn in the
structure favors the formation of spurious phases, mainly
Mn$_{3}$O$_{4}$\cite{joy}. Such a scenario is sustained by XRD
investigation of those samples, showing no variation in the LMO
XRD peaks positions (i.e., saturation in structural and/or
chemical composition of La-deficient LMO) and, more important, the
appearing of Mn$_{3}$O$_{4}$ diffraction peaks. On the contrary,
heavily La-deficient LMO thin films were proved to be structurally
stable when grown on suitable substrates\cite{gupta,apl2009}. In
such a specific form, XRD investigation shows a monotonic and
continuous change in structural lattice parameters, and no sign of
diffraction peaks associable to any spurious phase, mainly
Mn$_{3}$O$_{4}$. It is clear that the substrate plays a crucial
role in stabilizing the LMO structure, which otherwise would be no
longer stable for large La-deficiency.

\section{Transport properties}

Transport properties were also investigated as a function of the
La/Mn stoichiometry. Temperature dependence of the resistance for
a series of four LMO films, with different values of the La/Mn
ratio is reported in Fig.2 (data refer to the same LMO samples
which XRD structural characterization is reported in Fig.1).

\begin{figure}[!h]
\includegraphics[width=8.5 cm]{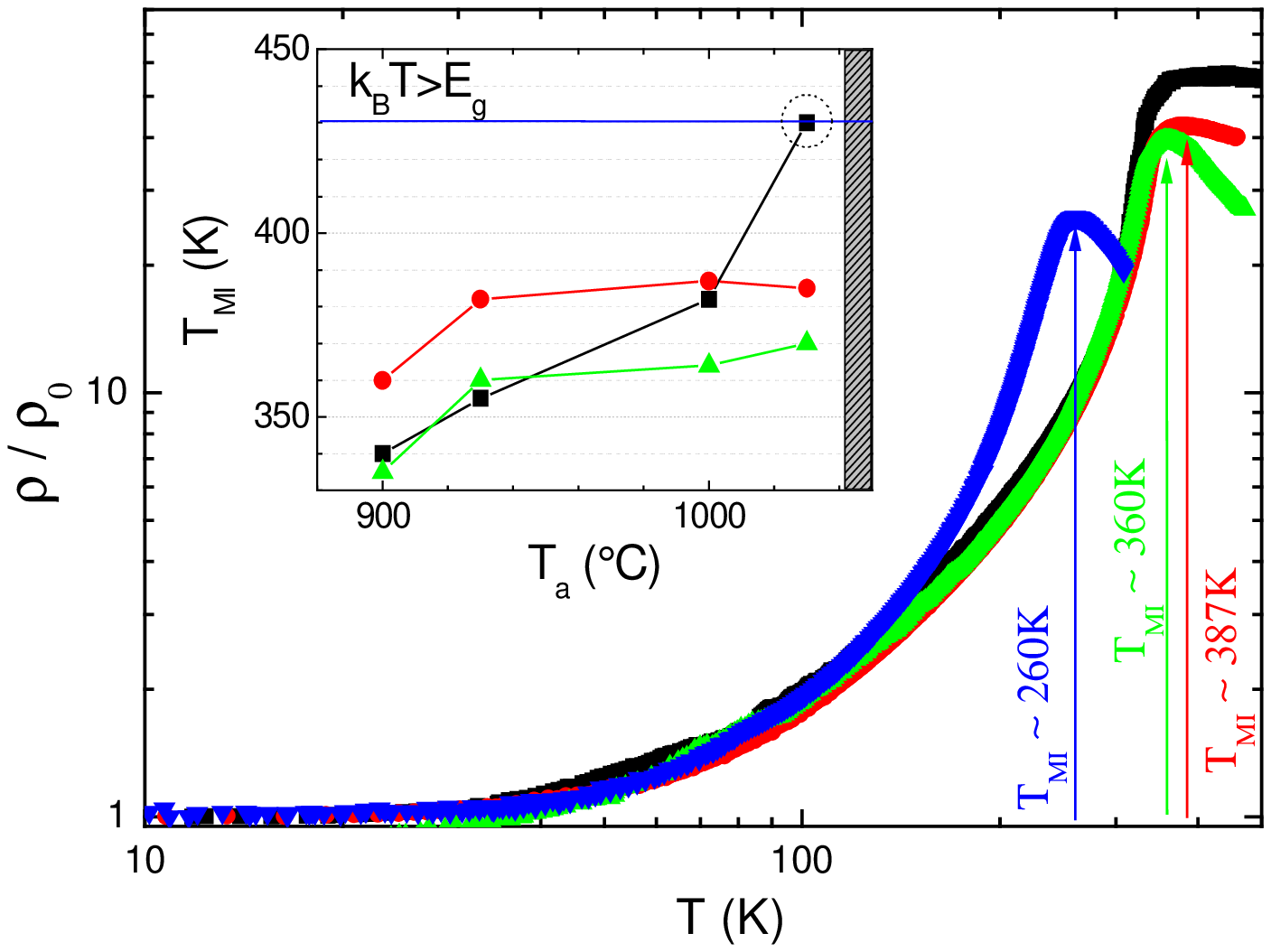}
\caption{\scriptsize{(Color online) Temperature dependence of the
resistivity (normalized to the residual resistivity) for a set of
four LMO samples with La/Mn ratio 0.66 (black), 0.88 (red), 0.98
(green) and 1.07 (blue), respectively (color code is the same than
figure 1). T$_{MI}$ are also indicated on the graph (T$_{MI}$ for
x = 0.66 sample is not indicated, being $\Delta \rho$/$\Delta$T
$<$ 0 at any temperature). In the inset, the metal-insulator
transition temperature T$_{MI}$ for different annealing process
(keeping fixed the duration at 54 hours and by changing the
temperature T$_{a}$) is also reported. Data refer to three set of
LMO samples, with La/Mn stoichiometric ratio of 0.66 (black
squares), 0.88 (red circles) and 0.98 (green triangles),
respectively. In the graph, the blue line corresponds to a value
of 430K.}}
\end{figure}

As the La-deficiency increases, more and more Mn ions are expected
to be pushed into a 4+ state. According to the general phase
diagram of manganites, T$_{MI}$ is expected to be maximum for a
33\% doping level of Mn$^{4+}$ and subsequently to drop to zero
for higher Mn$^{4+}$ concentration\cite{coey}. In
La$_{x}$MnO$_{3}$ where the Mn partial substitution at La-site
does not occur, such an optimal doping occurs for a 10\%
deficiency of La. A further decrease of the La-content should
force more a more Mn ions into a 4+ state, finally reaching a full
Mn$^{4+}$ population for a La/Mn ratio of 0.66. Mn$^{4+}$-doping
phase diagram is confirmed only for a small La-deficiency, in
which a monotonic increase of the T$_{MI}$ is indeed observed.
However, the highest metallicity ($\Delta\rho$/$\Delta$T $<$ 0 at
any temperature) is measured in La$_{0.66}$MnO$_{3-\delta}$, which
is otherwise expected to be insulating and antiferromagnetic.
Similarly, a monotonic increase of the Curie temperature (T$_{C}$)
has been measured within the LMO series, reaching the highest
value of 350K for the La$_{0.66}$MnO$_{3-\delta}$ sample. It is
evident that the metallic and ferromagnetic state in heavily
La-deficient manganite films can not be explained by the
self-doping mechanism.

We also investigated the possibility that the observed phenomena
could be due to a different oxygen diffusion into the various
samples. Such a diffusion can be affected by many extrinsic
factors such as chemical composition, strain, oxygen content, film
thickness, and so on. In order to rule out possible effects
related to the only oxygen content, we systematically investigated
the structural and the transport properties of our samples, by
changing the time and the temperature of a post-annealing process,
therefore probing different oxygen content in the LMO
films\cite{jap2007}. In the inset of Fig.2, the metal-insulator
transition temperature T$_{MI}$ for different post-annealing
process (keeping fixed the duration at 54 hours and by only
changing the temperature T$_{a}$) is reported. The temperature and
the duration of the annealing process directly affected the oxygen
content in LMO films, and, as a matter of fact, the T$_{MI}$ and
the resistivity values. As the annealing temperature increases,
the metallicity of the LMO samples also increases, thus confirming
the incorporation of more and more oxygen within the structure. In
this respect, by increasing the annealing temperature (i.e. oxygen
content) we are exploring the manganite T$_{MI}$ vs doping phase
diagram, which foresees both an under-doped and an over-doped
regime\cite{salamon,coey}. Nevertheless, if we look for instance
at the La/Mn=0.88 series, T$_{MI}$ show a maximum for an annealing
temperature of 1000$^{o}$C and subsequently a reduction of that
value for an annealing temperature of 1030$^{o}$C. Therefore, it
should be concluded that the La$_{0.88}$MnO$_{3-\delta}$ sample
annealed at 1000$^{o}$C is the optimally doped samples, showing
the optimal Mn$^{3+}$/Mn$^{4+}$ mixed population. However the
highest T$_{MI}$ among the La$_{0.66}$MnO$_{3-\delta}$ samples is
well above to such a value (in the inset of Fig.2, we have
indicated a value of T$_{MI}$ for this samples as high as 430K;
such a value corresponds to the destroying of any possible
polaronic insulating state by the thermal energy\cite{disorder}),
while it should have been equal to that measured in La/Mn=0.88
series, even if obtained with a different annealing process (i.e.
oxygen content). From these data it is clear that the physical
phenomena at play in heavily La-deficient LMO samples is not the
one at play in conventional manganites. Finally, it is well-known
that chemical and structural inhomogeneities (such as spurious
phases, oxygen stoichiometry, grain size, and others) can play an
important role in transport properties of manganite thin
films\cite{def1,def2,def3,arutalmo}. In this respect, as for
general inhomogeneities, strong structural and chemical disorder,
spurious insulating phase, impurities cooperate to worsen the
transport properties of manganite thin films (for instance,
because of low-resistance percolation paths, T$_{MI}$ usually
decreases and the resistivity increases). However, this is clearly
not what we observed, thus strongly supporting the hypothesis that
a novel mechanism is at play in heavily La-deficient thin films.

\section{X-ray Absorption Spectroscopy measurements}

To probe the electronic structure close to the Fermi level, which
is dominated by the Mn 3d and O 2p states, X-ray absorption
spectroscopy (XAS) at the Mn L$_{2,3}$ and O K edges were
performed at the ID08 beamline of ESRF in Grenoble. The anisotropy
effects related to the different doping levels were investigated
by varying the polarization of the grazing incidence synchrotron
radiation from horizontal (H-polarization) to vertical
(V-polarization) which correspond roughly to investigate the
out-of-plane and the in-plane directions, respectively. All Mn-L
spectra reported in Fig.3 show two broad multiplet structures,
L$_{3}$ and L$_{2}$, separated by spin-orbit splitting.

\begin{figure}[!h]
\includegraphics[width=8.5 cm]{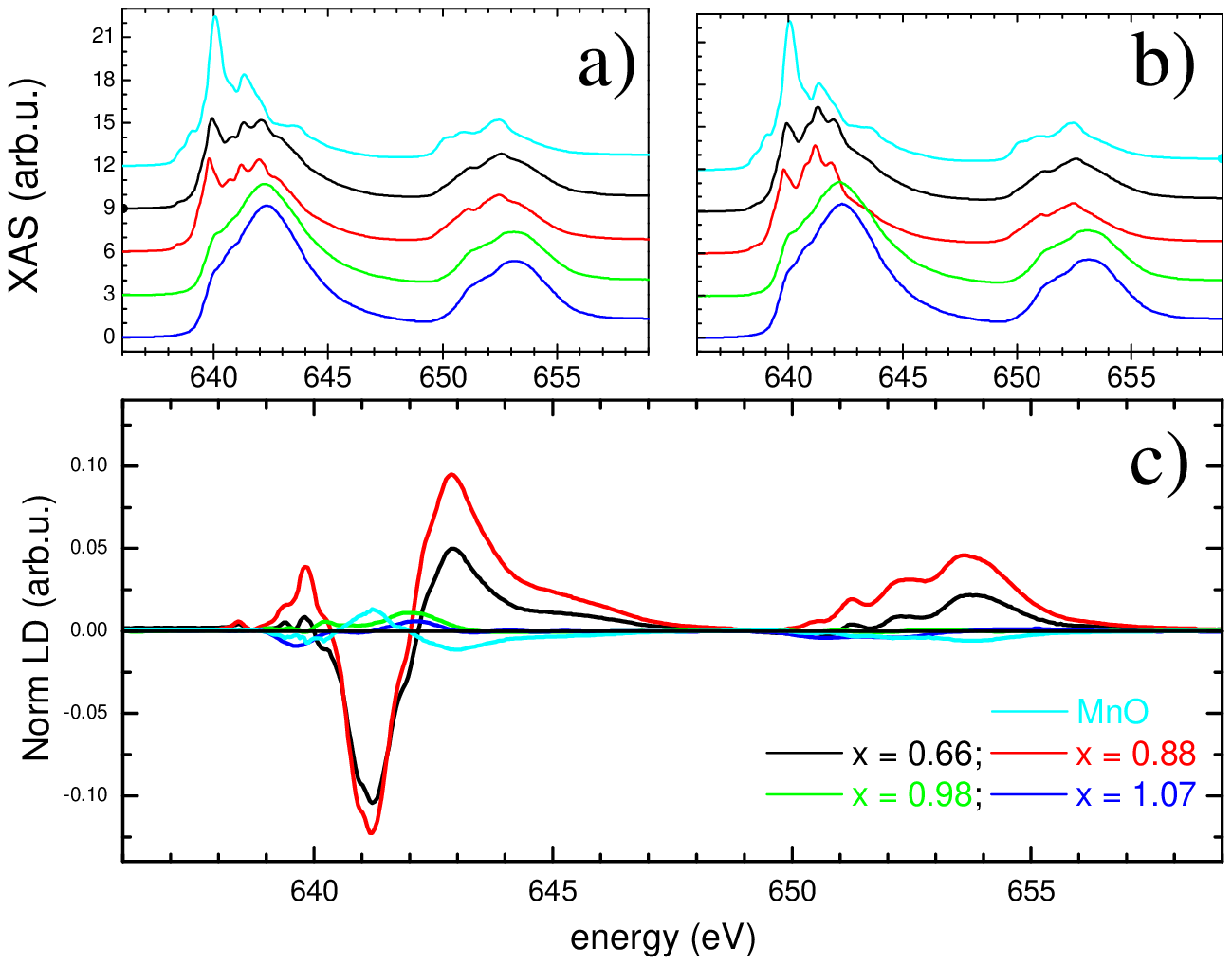}
\caption{\scriptsize{(Color online) Room temperature XAS (panel a,
in V-polarization and panel b in H-polarization), and LD at Mn
L$_{2,3}$ edge of LMO films with different La/Mn ratio. Data refer
to LMO film with La/Mn ratio 0.66 (black), 0.88 (red), 0.98
(green) and 1.07 (blue), respectively. XAS spectra for a MnO
single-crystal are also reported (cyan). All the spectra in V- and
H- polarization are shifted vertically for clarity.}}
\end{figure}

The effect of doping can be seen more clearly in the L$_{3}$
region where the spectra are similar to Mn$^{3+}$/Mn$^{4+}$
manganites when La/Mn$\gtrsim$1, while the increasing of Mn$^{2+}$
content in LMO films with La/Mn$<$1 gives rise to typical
pronounced features\cite{mitra}. The Linear Dichroism (LD)
obtained from the difference between the two XAS (V-H) is also
particularly interesting when the La/Mn$<$1. Indeed, the spectra
are similar to the theoretical calculations for the single
3d$^{4}$ configuration (Mn$^{3+}$) in case of a tetragonal
distortion of the octahedra, with an elongation of the
out-of-plane Mn-Mn distance\cite{aruta,aruta2}. Unlike Mn$^{3+}$,
it may be noted that Mn$^{4+}$ (3d$^{3}$) and Mn$^{2+}$ (3d$^{5}$)
do not tend to distort their own octahedral environment by
Jahn-Teller effect. Therefore, no-contribution is expected to the
orbital component of LD from Mn$^{4+}$ and Mn$^{2+}$. The
enhancement of LD intensity can be a consequence of the elongation
of the Mn$^{3+}$ octahedra to compensate the reduced O-Mn$^{2+}$-O
distances, thus being an indirect demonstration of the increased
Mn$^{2+}$ content. It is worthful to remark that in LaMnO$_{3}$
single-crystals Mn$^{2+}$-Mn$^{3+}$ coexistence was already
observed, pointing to an intrinsic mixed valence state in
insulating manganite, rather than Mn$^{2+}$-signal generated by
spurious MnO surface layers\cite{jimenez}. However, such
Mn$^{2+}$-Mn$^{3+}$ coexistence is expected to decrease by
increasing the metallicity of the system. In our investigation,
some films do not show any Mn$^{2+}$-features (ruling out the
possibility of a MnO surface layer due to post-annealing process
and/or surface reconstruction) and these last are observed in the
most metallic samples, while they are absent in the insulating
ones. Finally, XAS spectra at the Mn L-edge undoubtedly proves the
strict correlation between the La-deficiency and the presence of
Mn$^{2+}$ in heavily La-deficient LMO samples.

\section{Resonant Inelastic X-ray Spectroscopy measurements}

In order to fully comprehend the role of the Mn$^{2+}$ in the
transport mechanisms at play in La-deficient LMO films, it is
mandatory to unambiguously determine its structural configuration.
Manganites are mostly arranged in a perovskite structure, having
the general formula ABO$_{3}$, which can be seen as a stacking
sequence of AO- and BO$_{2}$-planes. Mn atoms are usually placed
at the B-site, while the RE$^{3+}$/AE$^{2+}$ atoms lie within the
AO-planes. In this respect, it is crucial to assign to the
Mn$^{2+}$-ions either the perovskite A-site or the B-site. In
order to do that, we exploited the almost unique capability of
L-edge RIXS to measure the set of the local $dd$ excitations to
assign the crystallographic site of divalent Mn$^{2+}$-ions and,
possibly, to derive some indications on its direct or indirect
involvement in the DE ferromagnetism in LMO films. The two
possible cation sites have different coordination and ion-oxygen
distances: site A (panel a in the inset in Fig.4) is
cubo-octahedral (12 O$^{2-}$ neighbors), with A-O distance $\simeq
2.7$ \AA, site B (panel b in the inset in Fig.4) is octahedral (6
O$^{2-}$ neighbors), with B-O distance $\simeq 1.9$ \AA. In a
simple point charge crystal field model one thus expects a much
smaller splitting of e$_{g}$ to t$_{2g}$ states at A than B sites.
Moreover the crystal field parameter $10Dq = E(e_{g})-E(t_{2g})$
should be negative at A and positive at B. As a reference we use
MnO, which has the NaCl structure: Mn$^{2+}$ ion is at octahedral
site with Mn-O distance $\simeq 2.2$ \AA, i.e., intermediate
between sites A and B in LMO. Naming $d$ the cation-ligand
distance, and recalling that point charge crystal
field\cite{figgis} predicts $10Dq$ $d^{-5}$, we expect $10Dq$ to
be smaller (greater) at site A (B) than in MnO. We have measured
the $dd$ excitation spectra of Mn$^{2+}$ in LMO and MnO using RIXS
excited at the Mn $L_3$ absorption edge. In particular we have
chosen the photon energy 640.0 eV, which is peculiar of Mn$^{2+}$
sites (see XAS spectra of Fig.3). The RIXS spectra are shown in
Fig.4.

\begin{figure}[!h]
\includegraphics[width=7 cm]{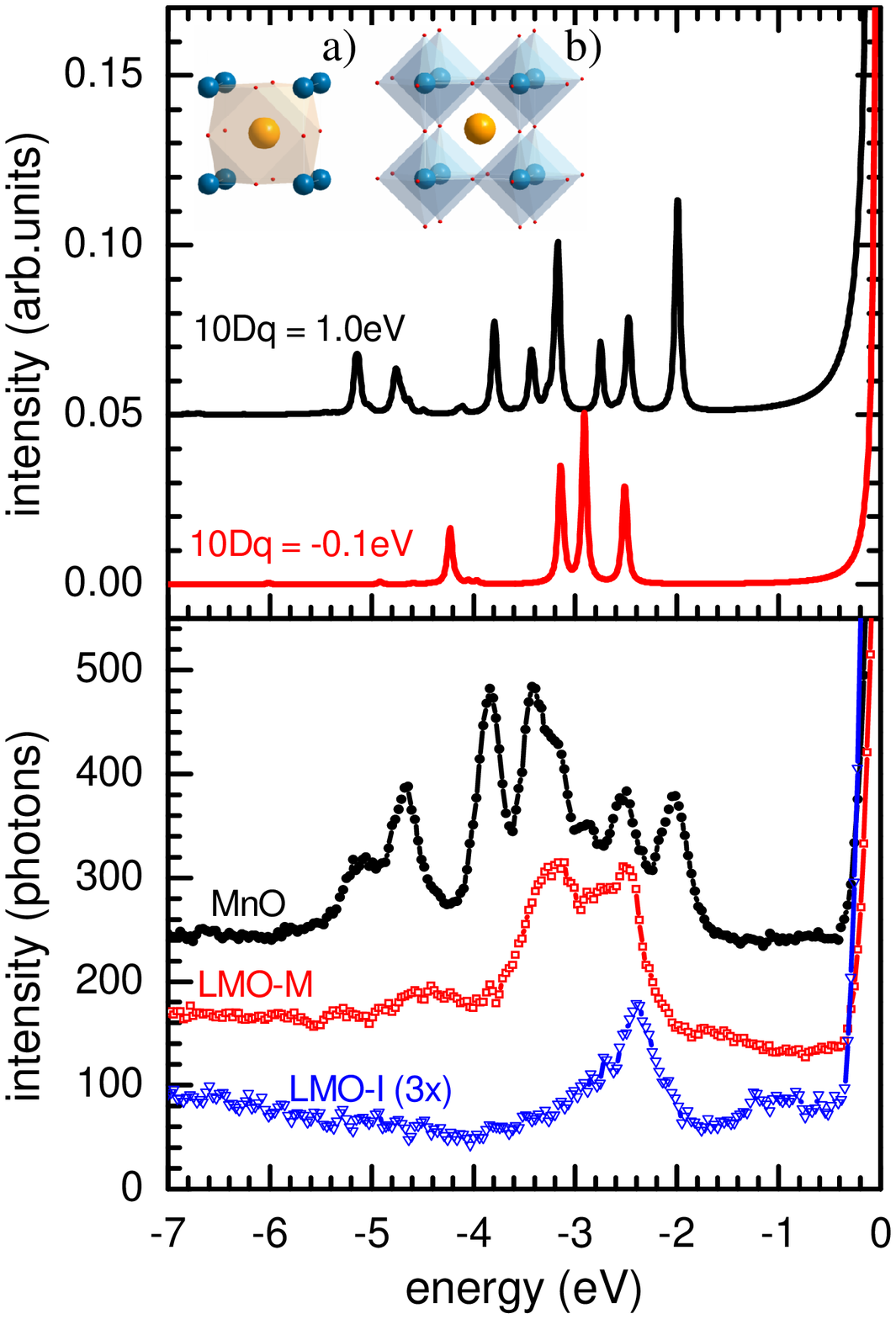}
\caption{\scriptsize{(Color online) (upper panel) Calculated RIXS
spectra within a single ion crystal field model and tuned 10Dq and
the rescaling of the inter-atomic Slater integrals; in the inset,
the oxygen coordination in undistorted perovskite around the
A-site (a) and the B-site (b) is also reported. (lower panel)
Experimental RIXS spectra for an insulating MnO single-crystal
(black), a Mn$^{2+}$-doped LMO metallic film (red, labelled LMO-M)
and an insulating LMO (blue, labelled LMO-I) thin films are
reported.}}
\end{figure}

In heavily La-deficient LMO metallic films, where XAS demonstrates
a strong presence of Mn$^{2+}$, the $dd$ excitations are very
different from those of MnO. To estimate quantitatively $10Dq$ we
have calculated the RIXS spectra within a single ion crystal field
model\cite{cowan,maurits} and tuned 10Dq and the rescaling of the
inter-atomic Slater integrals so to fit the main experimental
features. The result is shown in the top panel of Fig.4. For MnO
we essentially confirm the results of Ghiringhelli et
al\cite{giacomo_mno}: $10Dq = 1.0$ eV, Slater integrals rescaled
to 70\% of their Hartree-Fock value. For Mn$^{2+}$-doped LMO
films, we find a much smaller crystal field: $10Dq = -0.1$ eV,
Slater integrals rescaled to 64\%. This indicates univocally that
Mn$^{2+}$ sits at site A and that the Mn-O bond has a non
negligible covalent character (indicated by the strong
renormalization of Slater integrals). As further support to our
interpretation of the RIXS spectra, we show an example of
insulating LMO films, excited at the same energy as for the other
cases: the spectrum is incompatible with $dd$ excitations of
Mn$^{2+}$ and is rather typical of Mn$^{3+}$ in
magnanites\cite{giacomo_2009}.

\section{Theoretical model}

With the Mn$^{2+}$-ions indeed at the A-site, one would expect
that they could play the role of the divalent AE$^{2+}$-ion in
manganites by hole-doping the MnO$_{2}$ planes. In this respect,
the substitution of La$^{3+}$ ions with an atom having a smaller
ionic radius (namely 0.89 and 1.17\AA for Mn$^{2+}$ and La$^{3+}$,
respectively) should decrease the tolerance factor, thus pushing
the system into a more a more insulating state\cite{coey}. In this
respect, a novel approach based on electron-hole Bose liquid has
been recently proposed to describe metallic states in manganites
which are supposed to be in an insulating and antiferromagnetic
state. However, even though such phases are indeed metallic, they
are described in terms of "poor metallic phase" with respect to
the strong metallic and ferromagnetic state, which ultimately can
not described the enhanced metallicity that we observed in our
samples.

The observed enhanced metallicity can be understood by assuming
that the Mn$^{2+}$ ions at A-sites contribute to the electron
hopping via a new DE mechanism. The need for an alternative
conducting mechanism stems from the fact that a smaller tolerance
factor depresses the usual DE mechanism, but favor a new hopping
path involving Mn$^{2+}$ ions. Beside the hopping mechanism in the
traditional DE-framework (panel a of Fig.5), two extra possible
paths for the added hole were considered: namely, a hopping
mechanism between two Mn-ions directly through an intervening
Mn$^{2+}$-ion (panel b of Fig.5) and a \emph{Multiple}-DE hopping
path through the Oxygen-Mn-Oxygen ions (panel c of Fig.5).

\begin{figure}[!h]
\includegraphics[width=8.5 cm]{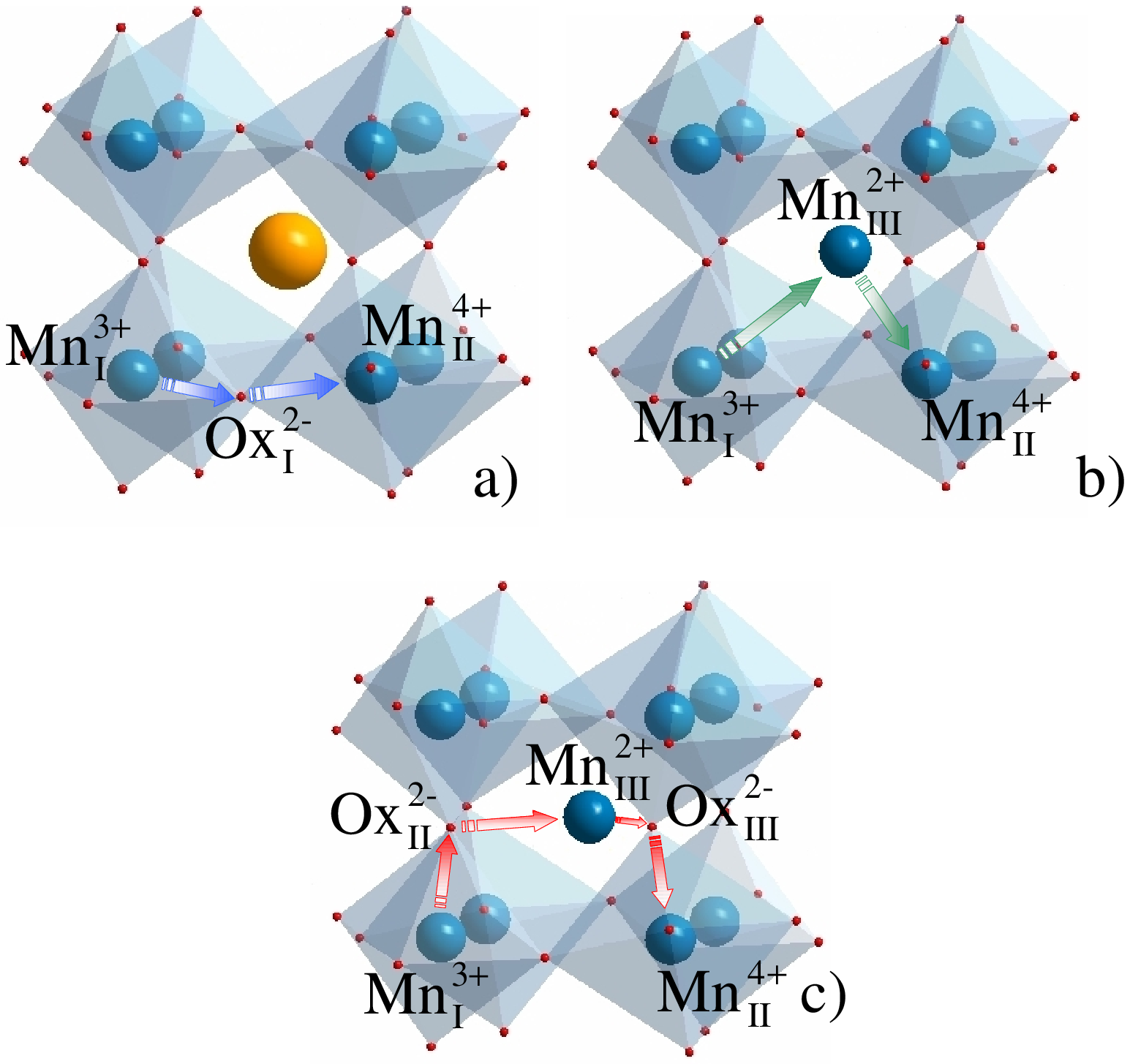}
\caption{\scriptsize{(Color online) Possible hopping mechanisms in
conventional and Mn$^{2+}$-substituted manganites. In panel a, the
traditional DE-mechanism (blue arrows). In panel b and c, two
possible additional hopping paths in Mn$^{2+}$-doped manganites
are reported: a direct hopping (panel b) between two Mn-ions
(Mn$_{I}$ and Mn$_{II}$, respectively) through intervening
Mn$^{2+}$-ion (namely, Mn$_{III}$) and a \emph{Multiple}-DE
hopping path (panel c) through the Oxygen-Mn-Oxygen ions
(Ox$_{II}$, Mn$_{III}$ and Ox$_{III}$, respectively).}}
\end{figure}

The effective hopping $t$ of Mn$^{3+}$-Mn$^{2+}$-Mn$^{4+}$ hopping
mechanism can be estimated following
Hasegawa-Anderson\cite{hasegawa} scheme, with the O$^{2-}$
replaced by Mn$^{2+}$ via $t=t_1^2/(E_1-E_2)$, where t$_{1}$ is
the resonance integral for the bond with the intermediate ion,
E$_{1}$ and E$_{2}$ are the energy levels of the two atomic
orbitals participating to the bond. In order to validate our
proposal we estimate the overall hopping of an e$_g$ electron
through Mn$^{2+}$ ions in an undistorted cubical lattice. We adopt
the values of ionic radii\cite{radii} and the energy
levels\cite{coey} that are reported in the literature. Using as
reference value the effective hopping Mn$^{3+}$-Mn$^{4+}$, through
the p$_{x}$ oxygen level along $x$ direction, t$_{x}$, we obtain
an effective hopping parameter of the order of 10$^{-3}$t$_{x}$
for both 3z$^{2}$-r$^{2}$ and x$^{2}$-y$^{2}$. This analysis shows
that a direct DE-hopping through Mn$^{2+}$ is very unlikely. On
the contrary, the effective DE hopping of the added hole through
the following path $Mn^{3+}-\,O^{2-}-\,Mn^{2+}-\,O^{2-}-\,Mn^{3+}$
is relevant. It can be estimated extending the approach used
above. By using a renormalization procedure that allows to
decimate the three intermediate sites\cite{pastore} we get:
\[
t_5=\frac{t_2^2t_1^2}{(E_1-E_2)(E_1^2-E_1(E_2+E_3)-2t_2^2+E_2E_3)}.
\]
where t$_{1}$ and t$_{2}$ are the resonance integrals between
Mn$^{4+}$-O$^{2-}$ and O$^{2-}$-Mn$^{2+}$ orbitals, respectively.
The energy levels E$_{1}$, E$_{2}$, E$_{3}$ are the energies of
the orbitals involved in the ions Mn$^{3+}$, O$^{2-}$, Mn$^{2+}$,
respectively. Assuming that the e$_{g}$ electrons are mainly in
the 3z$^{2}$-r$^{2}$ on Mn$^{3+}$ (based on XAS-LD analyses of
orbital ordering) and that the only possible states on Mn$^{2+}$
are xy and 3z$^{2}$-r$^{2}$ (due to orbital
symmetries\cite{slater}), our estimate is therefore
t$_{5}\simeq0.24$t$_{x}$.

Such a value has been obtained by assuming that all the ions are
structurally arranged in an undistorted perovskite. In fact,
manganites, and in particular LaMnO$_{3}$, generally show a
distorted orthorhombic/rhombohedral perovskite unit cell,
characterized by O-ions slightly displaced out of the joining
lines among the Mn-ions (see Fig.5). In this respect, because of
the buckling angle between the Mn-plane and the Mn-O bonds in
distorted perovskites, the superimposition of
Mn$_{I}$-Ox$_{I}$-Mn$_{II}$ orbitals is actually reduced, thus
subsequently reducing the t$_{x}$ value. Moreover, the smaller
size of Mn$^{2+}$-ionic radius (compared to La$^{3+}$'s ones)
should increase the cubic distortion of the perovskite cells, thus
further reducing the t$_{x}$ value. On the contrary, because of
the tilting of the oxygen octahedra, the distance between some
oxygen-ions and the central Mn$^{2+}$-ion should be shortened with
respect to the undistorted structure (in which the A-O distance is
$\sqrt{2}/2$ times unit cell lattice parameter). As a consequence,
the superimposition between O$^{2-}$-Mn$^{2+}$ orbitals actually
increases, thus increasing the t$_{5}$-value. Eventually, in real
distorted perovskite, a substantial increment of the
t$_{5}$/t$_{x}$-ratio is expected, thus pointing out towards a
significative role of the proposed new Multiple-DE conducting
path.

\section{Conclusions}

In conclusions, a new phenomenon for charge-hopping mechanism in
manganites (named \emph{Multiple} double-exchange, or
\emph{Multiple}-DE) has been observed. We have shown that, in
La$_{x}$MnO$_{3-\delta}$ manganite thin films, as La content
decrease, a partial substitution of Mn-ions at the La-site is
indeed possible. Polarization dependent XAS characterization has
demonstrated the relevant Mn$^{2+}$ content in heavily
La-deficient LMO samples. RIXS investigation has allowed us to
unambiguously assign the crystallographic site (namely, the
perovskite A-site) of the divalent Mn$^{2+}$. This last is
experimentally and theoretically proved to be more efficient with
respect to traditional DE-mechanism. Such a further increase of
the Curie temperatures in manganites could possible allow the use
of manganite-based devices at room temperatures, which nowadays is
limited because of the depression of magnetic alignment at that
specific temperatures. The very idea of using multiple-valence
atoms both as a dopant and as an active element electronically
involved in the conduction mechanism (namely, the Mn$^{2+}$ in
manganites) opens new scenarios in the study of physical
properties of CMR-materials. More important, such a strategy might
be also pursed in other strongly correlated electrons materials.

\begin{acknowledgments}
R.C.'s research activity has received funding from the European
Community's Seventh Framework Programme 2007-2011 under grant
agreement n.212348 NFFA.
\end{acknowledgments}

\end{document}